\documentclass[11pt]{article}

\usepackage{hyperref}
\usepackage[a4paper,text={6in,9in},centering]{geometry} % PAPER SIZE
\usepackage{amsmath}
\usepackage{amssymb}
\usepackage{amsbsy}
\usepackage{cite}
\makeatletter

\@addtoreset{equation}{section}
\makeatother

\title{
	\vskip -3em
	\begin{flushright}
	\small
	IFUP-TH/2010-45\\
	KOBE-TH-10-06
	\end{flushright}
	\vskip 5em
	\textbf{\large SUSY QM Meets 5d Gravity}
}
\author{
Satoshi Ohya\footnote{On leave of absence from Department of Physics, Kobe University, 1-1 Rokkodai, Nada, Kobe 657-8501, Japan.}\\[1ex]
\textit{\small INFN Sezione di Pisa, Largo Bruno Pontecorvo 3, 56127 Pisa, Italy}\\[1ex]
\texttt{\small email:\href{mailto:satoshi.ohya@pi.infn.it}{satoshi.ohya@pi.infn.it}}
}
\date{\small (Dated: December 3, 2010)}

\begin{document}
\maketitle
%%%%% ABSTRACT %%%%%
\begin{abstract}
We report hidden quantum mechanical supersymmetry structure in five-dimensional gravity with the Randall-Sundrum background.
We show that two $N=2$ supersymmetries are hidden in the spectrum.
\end{abstract}

\newpage
%%%%% SECTION 1 %%%%%
\section{Introduction} \label{sec:1}
Supersymmetric quantum mechanics (SUSY QM) provides a powerful tool to analyze spectrum of higher-dimensional field theory with extra compact dimensions.
In gauge theory with extra compact dimensions, gauge field that propagates in the bulk can be decomposed into normal modes with respect to extra dimensions and reduces to an infinite tower of massive vector Kaluza-Klein (KK) particles.
There, the longitudinal degrees of freedom of massive vector particles are provided by extra-dimensional component of higher-dimensional gauge field.
This mass generation mechanism is best described by hidden supersymmetry (SUSY) structure of higher-dimensional gauge theory \cite{Lim:2005rc}:
Normal modes of 4d and extra-dimensional components of gauge field have the same mass eigenvalues and form an $N=2$ quantum mechanical SUSY multiplet.
Extra-dimensional component, which is a 4d scalar, plays a role of would-be Nambu-Goldstone (NG) scalar boson and is smoothly absorbed into vector mode thanks to $N=2$ quantum mechanical SUSY \cite{Lim:2005rc}.
Schematically this SUSY structure can be described as follows:
\begin{align}
\text{gauge theory}:\,
\text{(vector mode)}
\overset{Q}{\underset{Q^{\dagger}}{\leftrightarrows}}
\text{(scalar mode)} \label{eq:1.1}
\end{align}
where $Q$ and $Q^{\dagger}$ are supercharge and its hermitian conjugate, which are the first-order differential operators with respect to the extra-dimensional coordinates.
As was discussed in Ref.\cite{Lim:2005rc} this $N=2$ quantum mechanical SUSY can be regarded as a remnant of higher-dimensional gauge symmetry and does not depend on any gauge choices.

In higher-dimensional gravity with extra compact dimensions, the mass generation mechanism for massive KK-graviton is best described by \textit{two} $N=2$ quantum mechanical SUSYs \cite{Lim:2007fy,Lim:2008hi}:
Normal modes of 4d graviton, vector and scalar components of metric are all degenerate and have the same mass eigenvalues.
Graviton and vector modes form an $N=2$ quantum mechanical SUSY multiplet, and vector and scalar modes form another SUSY multiplet.
Schematically speaking, these two $N=2$ SUSYs has the following structure:
\begin{align}
\text{gravity}:\,
\text{(graviton mode)}
\overset{\Tilde{Q}}{\underset{\Tilde{Q}^{\dagger}}{\leftrightarrows}}
\text{(vector mode)}
\overset{Q}{\underset{Q^{\dagger}}{\leftrightarrows}}
\text{(scalar mode)} \label{eq:1.2}
\end{align}
where $Q$ and $\Tilde{Q}$ are not necessarily the same differential operators.
Note also that $Q$ is not necessarily the same operator as gauge theory \eqref{eq:1.1}.
Just as in the case of gauge theory, two $N=2$ SUSYs can be regarded as a remnant of higher-dimensional general coordinate invariance and does not depend on any coordinate choices \cite{Lim:2007fy,Lim:2008hi}.

In this paper we would like to review our previous works on hidden SUSY structure of 5d gravity.
For the sake of simplicity we concentrate ourselves to pure gravitational fluctuations on background geometry described by the Randall-Sundrum metric \cite{Randall:1999ee,Randall:1999vf}.
The rest of this paper is organized as follows.
In section \ref{sec:2} we set up the model and study gauge-fixed action to quadratic order.
In section \ref{sec:3} we show that two $N=2$ SUSYs are hidden in the 4d mass spectrum.
In section \ref{sec:4} we study allowed boundary conditions consistent with two $N=2$ SUSYs.
We will see that two SUSYs severely restrict the possible boundary conditions for metric fluctuations and show that allowed boundary condition is uniquely determined.
Section \ref{sec:5} is devoted to conclusions and some speculations.

%%%%% SECTION 2 %%%%%
\section{Gauge-fixed action to quadratic order} \label{sec:2}
In this section we study the bulk gravity action up to quadratic order of metric fluctuations around the background Randall-Sundrum metric with a gauge-fixing term.
The purpose of this section is to identify the squared-mass operators for 4d graviton mode, vector mode (would-be NG vector boson) and scalar mode (would-be NG scalar boson).

To this end let us start with the Einstein-Hilbert action in the bulk
\begin{align}
S_{\text{EH}}
&= 	\int\!\mathrm{d}^{4}x\int_{z_{1}}^{z_{2}}\!\!\!\mathrm{d}z\sqrt{-G}
	\bigl(M^{3}R - \Lambda\bigr), \label{eq:2.1}
\end{align}
where $M$ is the mass scale of 5d gravity and $\Lambda<0$ is the bulk cosmological constant.
We study gravitational fluctuations around the Randall-Sundrum metric \cite{Randall:1999ee,Randall:1999vf} in the conformal coordinate
\begin{align}
G_{MN}(x, z)
&= 	\mathrm{e}^{2A(z)}\bigl[\eta_{MN} + h_{MN}(x, z)\bigr], \quad
A(z)
= 	-\log\frac{z}{R}, \label{eq:2.2}
\end{align}
where $R$ is the AdS radius given by $R = \sqrt{12M^{3}/(-\Lambda)}$ and $h_{MN}$ is the metric fluctuation.
The location of UV brane $z=z_{1}$ is chosen as $z_{1} = R$ in order to make the background metric be Minkowski metric $\eta_{MN}$ on the UV brane.
The location of IR brane $z=z_{2}$ is left arbitrary.

For the following discussions the most useful parameterization of $h_{MN}$ is turned out to be of the form
\begin{align}
h_{MN}
&= 	\begin{pmatrix}
	h_{\mu\nu} - \frac{1}{2}\eta_{\mu\nu}\phi 	& h_{\mu 5} \\
	h_{\mu 5} 							& \phi
	\end{pmatrix}, \label{eq:2.3}
\end{align}
which enables us to identify the spectrum of 5d gravity.

Next we wish to fix the general coordinate invariance by adding a gauge-fixing action.
Just like an ordinary spontaneously broken gauge theory, we want to use one-parameter family of gauge-fixing functions which completely fix the general coordinate invariance and further remove the unwanted quadratic mixings among $h_{\mu\nu}$, $h_{\mu5}$ and $\phi$.
In Ref.\cite{Lim:2008hi} such a one-parameter family of gauge choices (i.e. $R_{\xi}$-gauge) for 5d gravity has been proposed and studied in great detail.
To quadratic order of metric fluctuations such gauge-fixing action is given as follows
\begin{align}
S_{\text{GF}}
&= 	M^{3}\int\!\mathrm{d}^{4}x\int_{z_{1}}^{z_{2}}\!\!\!\mathrm{d}z\,
	\mathrm{e}^{3A(z)}
	\left[
	-\frac{1}{2\xi}\bigl(F_{\mu}[h]\bigr)^{2}
	-\frac{1}{2\xi}\bigl(F_{5}[h]\bigr)^{2}
	\right], \label{eq:2.4}
\end{align}
where $F_{\mu}$ and $F_{5}$ are the gauge-fixing functions defined by
\begin{subequations}
\begin{align}
F_{\mu}[h]
&:= 	- \partial^{\lambda}h_{\lambda\mu}
	+ \frac{1}{2}\left(2 - \frac{1}{\xi}\right)\partial_{\mu}h
	- \xi(\partial_{z} + 3A^{\prime})h_{\mu5}, \label{eq:2.5a}\\
F_{5}[h]
&:= 	\frac{1}{2}\partial_{z}h
	- \partial^{\mu}h_{\mu5}
	- \frac{3\xi}{2}(\partial_{z} + 2A^{\prime})\phi, \label{eq:2.5b}
\end{align}
\end{subequations}
with $\xi$ being a real parameter that ranges from $-\infty$ to $+\infty$.
The limit $\xi \to \infty$ corresponds to the unitary gauge \cite{Lim:2008hi}.
For the sake of simplicity, however, in the following discussions we set $\xi=1$, which is an analogue of 't Hooft-Feynman gauge in ordinary spontaneously broken gauge theory.
In this gauge the bulk gauge-fixed action takes the following simple quadratic form
\begin{align}
S_{\text{EH}} + S_{\text{GF}}^{(\xi=1)}
&= 	M^{3}\int\!\mathrm{d}^{4}x\int_{z_{1}}^{z_{2}}\!\!\!\mathrm{d}z\,
 	\biggl\{\frac{1}{4}
	\Bar{h}^{\mu\nu}
	\left[
	\frac{1}{2}(\eta_{\mu\rho}\eta_{\nu\sigma} + \eta_{\mu\sigma}\eta_{\nu\rho} - \eta_{\mu\nu}\eta_{\rho\sigma})
	\left(
	\Box - H_{2}
	\right)
	\right]
	\Bar{h}^{\rho\sigma} \nonumber\\
& 	\hspace{8em}
	+\frac{1}{2}
	\Bar{h}^{\mu 5}
	\left[
	\eta_{\mu\nu}
	\left(
	\Box - H_{1}
	\right)
	\right]
	\Bar{h}^{\nu 5} \nonumber\\
& 	\hspace{8em}
	+\frac{3}{8}
	\Bar{\phi}
	\left[
	\Box - H_{0}
	\right]
	\Bar{\phi}
	\biggr\} + O(\Bar{h}^{3}), \label{eq:2.6}
\end{align}
where $\Box = \partial_{\mu}\partial^{\mu}$.
In Eq.\eqref{eq:2.6} we have redefined the metric fluctuation as $\Bar{h}_{MN}(x, z) := \mathrm{e}^{(3/2)A(z)}h_{MN}(x, z)$, which enables us to remove the first-order derivative terms with respect to $z$ from the quadratic action.
$H_{s}$ ($s=0,1,2$) is the second-order differential operator (mass operator) which does not contain first-order derivative thanks to the redefinition of the fluctuations and takes the following Schr\"odinger form
\begin{align}
H_{s}
&= 	-\partial_{z}^{2} + \frac{s^{2} - 1/4}{z^{2}}, \quad
s=0,1,2. \label{eq:2.7}
\end{align}
The problem to find the spectrum of 5d gravity is thus reduced to the eigenvalue problem of ordinary time-independent Schr\"odinger equation $H_{s}f_{s}(z) = m^{2}f_{s}(z)$, where $f_{s}(z)$ is a square-integrable function on an interval $(z_{1}, z_{2})$.
As mentioned in section \ref{sec:1}, the spectrum of $H_{0}$, $H_{1}$ and $H_{2}$ must be all degenerate (up to possible zero-modes) otherwise cancellations between unphysical degrees of freedom would become incomplete and hence spurious massive vector and scalar modes could contribute to the physical amplitudes when one computes amplitudes of physical processes.
As we will see in the subsequent sections, the spectra of these three Hamiltonians are indeed degenerate. This three-fold degeneracy is guaranteed by two $N=2$ quantum mechanical SUSYs.

%%%%% SECTION 3 %%%%%
\section{SUSY in the spectrum} \label{sec:3}
In this section we reveal the hierarchical structure of Hamiltonians $H_{0}$, $H_{1}$ and $H_{2}$, and show that two $N=2$ quantum mechanical SUSYs are hidden in the spectrum.
To this end we first note that the Hamiltonian $H_{s}$ can be factorized as follows
\begin{align}
H_{s}
&= 	\mathcal{A}_{s}^{\dagger}\mathcal{A}_{s}, \label{eq:3.1}
\end{align}
where $\mathcal{A}_{s}$ and $\mathcal{A}^{\dagger}_{s}$ are the first-order differential operators defined by
\begin{align}
\mathcal{A}_{s}
&:= 	+ \partial_{z} + \left(s+\frac{1}{2}\right)A^{\prime}(z), \quad
\mathcal{A}^{\dagger}_{s}
:= 	- \partial_{z} + \left(s+\frac{1}{2}\right)A^{\prime}(z). \label{eq:3.2}
\end{align}
Prime ($\prime$) indicates the derivative with respect to $z$.
This factorization of Hamiltonian is crucial for our discussion and indeed the most important ingredient of SUSY QM (see for review \cite{Cooper:1994eh}).
It should be noted that the warp factor $A(z)$ plays a role of superpotential (prepotential) in SUSY QM.
Before going to discuss the hidden SUSY structure of the theory, we would like to mention about the following two points.
The first is that factorization of Hamiltonian is always possible for any Hamiltonian $H = -\partial_{z}^{2} + V(z)$ as $H = \mathcal{A}^{\dagger}\mathcal{A}$ with the definitions
\begin{align}
\mathcal{A}
&:= 	+(\text{zero-mode})\partial_{z}\frac{1}{(\text{zero-mode})}, \quad
\mathcal{A}^{\dagger}
:= 	-\frac{1}{(\text{zero-mode})}\partial_{z}(\text{zero-mode}), \label{eq:3.3}
\end{align}
where ``$(\text{zero-mode})$'' indicates the zero-eigenvalue solution to the Schr\"odinger equation.\footnote{If $H$ does not have zero-eigenvalue solution, subtract the ground state energy from $H$ and then use the ground state solution instead of zero-mode.}
Indeed,  one can easily check that $H_{s}f_{s}(z) = m^{2}f_{s}(z)$ has the zero-eigenvalue solution $\mathrm{e}^{-(s+1/2)A(z)}$ such that $\mathcal{A}_{s}$ and $\mathcal{A}_{s}^{\dagger}$ can be written as $\mathcal{A}_{s} = + \mathrm{e}^{-(s+1/2)A(z)}\partial_{z}\mathrm{e}^{+(s+1/2)A(z)}$ and $\mathcal{A}_{s}^{\dagger} = - \mathrm{e}^{+(s+1/2)A(z)}\partial_{z}\mathrm{e}^{-(s+1/2)A(z)}$.
(Note that this discussion does not depend on whether zero-mode indeed appears in the spectrum or not.)

The second point we would like to mention here is that, because $H_{s}$ is the second-order differential operator, we have two independent zero-eigenvalue solutions, which implies that $H_{s}$ can be factorized into two different ways.
Noting that the Hamiltonian \eqref{eq:2.7} possesses the $\mathbb{Z}_{2}$ symmetry $s \to -s$, we immediately see that there exists another zero-eigenvalue solution $\mathrm{e}^{-(-s+1/2)A(z)}$ and the corresponding factorization $H_{s} = \mathcal{A}_{-s}^{\dagger}\mathcal{A}_{-s} = \mathcal{A}_{s-1}\mathcal{A}_{s-1}^{\dagger}$, where the second equality follows from the identities $\mathcal{A}_{-s} = -\mathcal{A}_{s-1}^{\dagger}$ and $\mathcal{A}_{-s}^{\dagger} = -\mathcal{A}_{s-1}$.\footnote{
This discussion cannot be applied to the case $s=0$ (scalar sector) because $s=0$ is the fixed point of $\mathbb{Z}_{2}$-transformation $s \to -s$.
However, refactorization of Hamiltonian $H_{0}$ is possible by using the other zero-eigenvalue solution $A(z)\mathrm{e}^{-(1/2)A(z)}$, which contains logarithm of $z$ and hence non-polynomial.
}
In the end, $H_{s}$ admits the following two different factorizations
\begin{align}
H_{s}
&= 	\mathcal{A}_{s-1}\mathcal{A}_{s-1}^{\dagger}
= 	\mathcal{A}_{s}^{\dagger}\mathcal{A}_{s}. \label{eq:3.4}
\end{align}
We note that the second equality of \eqref{eq:3.4}, which we call refactorization of Hamiltonian, is just achieved by changing the ordering of $\mathcal{A}_{s-1}$ and $\mathcal{A}_{s-1}^{\dagger}$ and shifting the parameter $s \to s+1$.
This is nothing but the consequence of shape invariance nature of inverse square potential (see for example \cite{Cooper:1994eh}).
In gravity language, on the other hand, this refactorization of Hamiltonian is thanks to the relation $(A^{\prime}(z))^{2} - A^{\prime\prime}(z) = 0$, which is one of background Einstein equations.
If the warp factor $A(z)$ does not satisfy the background Einstein equations, it turns out that there is no SUSY structure in the spectrum and the three-fold degeneracy among graviton, vector and scalar modes disappears.

Now, by making use of the relation \eqref{eq:3.4} we obtain the hierarchy of Hamiltonians
\begin{alignat}{7}
\text{scalar sector}:~
&H_{0} = \mathcal{A}^{\dagger}_{0}\mathcal{A}_{0}
&
&
&
&
&
&\nonumber\\
\text{vector sector}:~
&H_{1} = \mathcal{A}_{0}\mathcal{A}^{\dagger}_{0}~
&
&= \mathcal{A}^{\dagger}_{1}\mathcal{A}_{1}
&
&
&
&\nonumber\\
\text{graviton sector}:~
&H_{2} =
&
&= \mathcal{A}_{1}\mathcal{A}^{\dagger}_{1}~
&= \mathcal{A}_{2}^{\dagger}\mathcal{A}_{2}
&
&\nonumber\\
&H_{3} =
&
&
&= \mathcal{A}_{2}\mathcal{A}^{\dagger}_{2}
&= \cdots
&\nonumber\\
&~~\vdots
&
&
&
&\,\,\,\vdots
& \nonumber
\end{alignat}
In 5d gravity there appear the first three lines of hierarchy in the spectrum.
Naively this hierarchy of Hamiltonians can be extended infinitely.
Therefore, one might expect that if we consider 5d massless spin-$N$ field theory with the Randall-Sundrum background, we would obtain the first $N$ lines of this type of hierarchy in the spectrum.
In this paper we will not study higher-spin field theory and not discuss this point further, however, we have to emphasize that the above argument is valid only in the bulk.
Whether the eigenvalues of these Hamiltonians are indeed degenerate or not strongly depends on the boundary conditions.
As we will see in the next section, we can show that in our setting it is impossible to construct a hierarchy of isospectral Hamiltonians beyond three-fold degeneracy.
Before going to discuss this point further, let us turn to the analysis of hidden SUSY structure of 5d gravity.

%%%%% SECTION 3.1 %%%%%
\subsection{SUSY between scalar \& vector sectors (vector \& graviton sectors)}
Now it is obvious that there exists an $N=2$ SUSY structure between the scalar and vector sectors.
Indeed, by introducing the differential operators
\begin{align}
H
&= 	\begin{pmatrix}
	H_{0} 	& 0 \\
	0 		& H_{1}
	\end{pmatrix}, \quad
Q
= 	\begin{pmatrix}
	0 				& 0 \\
	\mathcal{A}_{0} 	& 0
	\end{pmatrix}, \quad
Q^{\dagger}
= 	\begin{pmatrix}
	0 	& \mathcal{A}_{0}^{\dagger} \\
	0 	& 0
	\end{pmatrix}, \label{eq:3.5}
\end{align}
which act on the two-component vector $(f_{0}(z), f_{1}(z))^{T}$ ($T$ stands for transposition), we have the $N=2$ SUSY algebra\footnote{
$N=2$ SUSY will become clearer by taking the hermitian linear combinations $Q_{1} = Q + Q^{\dagger}$ and $Q_{2} = i\sigma_{3}Q_{1}$ ($\sigma_{3} = \left(\begin{smallmatrix}1 & 0\\ 0 & -1 \end{smallmatrix}\right)$: $\mathbb{Z}_{2}$-grading operator), which satisfy the standard $N=2$ SUSY algebra
\begin{align}
\{Q_{i}, Q_{j}\}
= 	2\delta_{ij}H, \quad
[H, Q_{i}]
= 	[H, \sigma_{3}]
= 	0, \quad
i,j=1,2. \nonumber
\end{align}
}
\begin{align}
& 	\{Q, Q^{\dagger}\}
= 	H, \quad
	\{Q, Q\}
= 	\{Q^{\dagger}, Q^{\dagger}\}
= 	0, \quad
 	[H, Q]
= 	[H, Q^{\dagger}]
= 	0. \label{eq:3.6}
\end{align}
SUSY relations between $f_{0}$ and $f_{1}$ are
\begin{align}
Q
\begin{pmatrix}
f_{0}(z) \\
0
\end{pmatrix}
= 	m
	\begin{pmatrix}
	0 \\
	f_{1}(z)
	\end{pmatrix}
\quad\text{and}\quad
Q^{\dagger}
\begin{pmatrix}
0 \\
f_{1}(z)
\end{pmatrix}
= 	m
	\begin{pmatrix}
	f_{0}(z) \\
	0
	\end{pmatrix}. \label{eq:3.7}
\end{align}
Thus we can say that scalar mode $f_{0}$ and vector mode $f_{1}$ form an $N=2$ SUSY multiplet.
Similarly, we have another $N=2$ SUSY structure between vector and graviton sectors.
By introducing the differential operators $\Tilde{H} = \mathrm{diag}(H_{1}, H_{2})$, $\Tilde{Q} = \left(\begin{smallmatrix} 0 & 0\\ \mathcal{A}_{1} & 0\end{smallmatrix}\right)$ and $\Tilde{Q}^{\dagger} = \left(\begin{smallmatrix} 0 & \mathcal{A}_{1}^{\dagger}\\ 0 & 0\end{smallmatrix}\right)$, which act on the two-component vector $(f_{1}(z), f_{2}(z))^{T}$, we have the same $N=2$ SUSY algebra as \eqref{eq:3.6}.

%%%%% SECTION 3.2 %%%%%
\subsection{SUSY between scalar \& graviton sectors}
An interesting point to note is that there appears a nonlinear SUSY structure between the scalar and graviton sectors.
Indeed, by introducing the differential operators
\begin{align}
\Hat{H}
&= 	\begin{pmatrix}
	H_{0} 	& 0 \\
	0 		& H_{2}
	\end{pmatrix}, \quad
\Hat{Q}
= 	\begin{pmatrix}
	0 							& 0 \\
	\mathcal{A}_{1}\mathcal{A}_{0} 	& 0
	\end{pmatrix}, \quad
\Hat{Q}^{\dagger}
= 	\begin{pmatrix}
	0 	& \mathcal{A}_{0}^{\dagger}\mathcal{A}_{1}^{\dagger} \\
	0 	& 0
	\end{pmatrix}, \label{eq:3.8}
\end{align}
which act on the two-component vector $(f_{0}(z), f_{2}(z))^{T}$, we have the \textit{nonlinear} algebra
\begin{align}
& 	\{\Hat{Q}, \Hat{Q}^{\dagger}\}
= 	\Hat{H}^{2}, \quad
 	\{\Hat{Q}, \Hat{Q}\}
= 	\{\Hat{Q}^{\dagger}, \Hat{Q}^{\dagger}\}
= 	0, \quad
 	[\Hat{H}, \Hat{Q}]
= 	[\Hat{H}, \Hat{Q}^{\dagger}]
= 	0. \label{eq:3.9}
\end{align}
This is one of the simplest nonlinear extensions of $N=2$ SUSY discussed in the literature under the name of the second-order derivative SUSY \cite{Andrianov:1994aj,Andrianov:1995xt,FernandezC:1996hh,Andrianov:2003dg} or $\mathcal{N}$-fold SUSY with $\mathcal{N}=2$ \cite{Aoyama:2001ca}.

\begin{figure}[t]
\begin{center}
\setlength\unitlength{1em}
\begin{picture}(33,2)(0,-0.7)
\put(0,0){(graviton mode)}
\put(13.5,0){(vector mode)}
\put(26.25,0){(scalar mode)}

\put(7.5,-0.7){\small $N=2$ SUSY}
\put(10.25,0.3){\vector(1,0){2.5}}
\put(10.25,0.3){\vector(-1,0){2.5}}

\put(20.25,-0.7){\small $N=2$ SUSY}
\put(23,0.3){\vector(1,0){2.5}}
\put(23,0.3){\vector(-1,0){2.5}}

\put(11.5,1.8){\small 2nd order derivative SUSY}
\put(4,1.5){\line(1,0){25}}
\put(4,1.5){\vector(0,-1){0.8}}
\put(29,1.5){\vector(0,-1){0.8}}
\end{picture}
\caption{Hidden quantum mechanical SUSY structure in 5d gravity.}
\label{fig:1}
\end{center}
\end{figure}
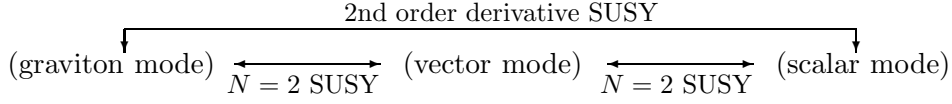

We summarize the hidden SUSY structure in the spectrum of 5d gravity in Figure \ref{fig:1}.

%%%%% SECTION 4 %%%%%
\section{Boundary conditions and SUSY} \label{sec:4}
So far we have seen that two $N=2$ SUSYs are hidden in the mass eigenvalue problem of 5d gravity.
However, as mentioned before, whether mass eigenvalues are indeed degenerate or not strongly depends on the boundary conditions.
In this section we study boundary conditions consistent with the hermiticity of each Hamiltonian and two $N=2$ SUSYs.

To begin with, let us first recall the most general boundary conditions consistent with the hermiticity of $H_{0} = \mathcal{A}_{0}^{\dagger}\mathcal{A}_{0}$ with respect to the inner product $\langle f_{0}|g_{0}\rangle = \int_{z_{1}}^{z_{2}}\mathrm{d}z\,f_{0}^{\ast}(z)g_{0}(z)$.
As discussed in Ref.\cite{Nagasawa:2008an} hermiticity requirement $\langle f_{0}|H_{0}g_{0}\rangle - \langle H_{0}f_{0}|g_{0}\rangle = 0$ leads to the following $U(1) \times U(1)$ parameter family of boundary conditions
\begin{align}
f_{0}(z_{i}) + L_{0}\cot\left(\frac{\theta_{i}}{2}\right)(\mathcal{A}_{0}f_{0})(z_{i})
&= 	0, \quad
0\leq\theta_{i}<2\pi, \quad
i,j=1,2, \label{eq:4.1}
\end{align}
where $(\theta_{1}, \theta_{2})$ are two independent parameters of the group $U(1) \times U(1)$, and $L_{0}$ is an arbitrary length scale which is just introduced to adjust the length dimension of the equation \eqref{eq:4.1}.
Next we use the SUSY relations $\mathcal{A}_{0}f_{0}(z) = mf_{1}(z)$ and $\mathcal{A}_{0}^{\dagger}f_{1}(z) = mf_{0}(z)$.
These relations lead to the following eigenvalue dependent boundary condition
\begin{align}
(\mathcal{A}_{0}^{\dagger}f_{1})(z_{i}) + m^{2}L_{0}\cot\left(\frac{\theta_{i}}{2}\right)f_{1}(z_{i})
&= 	0, \quad
0\leq\theta_{i}<2\pi, \quad
i,j=1,2. \label{eq:4.2}
\end{align}
Because boundary condition should not depend on the eigenvalue (otherwise superposition of distinct eigenmodes could not have any definite boundary behavior and hence would become meaningless), we can conclude that hermiticity and SUSY become compatible if and only if $\theta_{i} = 0$ or $\pi$.
Thus the boundary condition consistent with $N=2$ SUSY is characterized by the group $\mathbb{Z}_{2} \times \mathbb{Z}_{2} \subset U(1) \times U(1)$ \cite{Nagasawa:2008an}.
Consequently, at $z=z_{i}$ we have only two choices
\begin{subequations}
\begin{alignat}{4}
&\text{i})~
&\theta_{i} = 0: \quad
&(\mathcal{A}_{0}f_{0})(z_{i}) = 0
&\quad\&\quad
&f_{1}(z_{i}) = 0,&\label{eq:4.3a}\\
&\text{ii})~
&\theta_{i} = \pi: \quad
&f_{0}(z_{i}) = 0
&\quad\&\quad
&(\mathcal{A}_{0}^{\dagger}f_{1})(z_{i}) = 0.& \label{eq:4.3b}
\end{alignat}
\end{subequations}
Similarly, if we start from the requirement of the hermiticity of $H_{1} = \mathcal{A}_{1}^{\dagger}\mathcal{A}_{1}$ and then use the SUSY relations between $f_{1}$ and $f_{2}$, we have the following boundary conditions:
\begin{subequations}
\begin{alignat}{4}
&\text{i})~
&\theta_{i} = 0: \quad
&(\mathcal{A}_{1}f_{1})(z_{i}) = 0
&\quad\&\quad
&f_{2}(z_{i}) = 0,&\label{eq:4.4a}\\
&\text{ii})~
&\theta_{i} = \pi: \quad
&f_{1}(z_{i}) = 0
&\quad\&\quad
&(\mathcal{A}_{1}^{\dagger}f_{2})(z_{i}) = 0.&\label{eq:4.4b}
\end{alignat}
\end{subequations}
In order to respect two $N=2$ SUSYs we have to impose boundary conditions consistent with Eqs.\eqref{eq:4.3a}--\eqref{eq:4.4b}.
Since any two of the conditions $f_{1}(z_{i}) = 0$, $(\mathcal{A}_{0}^{\dagger}f_{1})(z_{i}) = 0$ and $(\mathcal{A}_{1}f_{1})(z_{i}) = 0$ cannot be compatible with each other,\footnote{
Exceptional case can arise if the derivative of warp factor diverges at the boundary $A^{\prime}(z) \stackrel{z\to z_{i}}{\to} \infty$, which is not the case of the Randall-Sundrum background $A(z) = -\log(z/R)$ because $z_{1} \neq 0$.
(On the contrary, one can show that there exists a set of boundary conditions consistent with an infinite tower of $N=2$ SUSYs at the AdS boundary $z=0$.)
The simplest quantum mechanical example of such exceptions is the hierarchy of Hamiltonians obtained by starting from the system of infinitely deep well potential.
In this case we can construct a hierarchy of isospectral Hamiltonians beyond three-fold degeneracy.
} such boundary condition is uniquely determined:
\begin{align}
(\mathcal{A}_{0}f_{0})(z_{i})
= 	0 \quad\&\quad
f_{1}(z_{i})
= 	0 \quad\&\quad
(\mathcal{A}_{1}^{\dagger}f_{2})(z_{i})
= 	0, \quad i=1,2. \label{eq:4.5}
\end{align}
Other choices of boundary conditions cannot be consistent with two $N=2$ SUSYs and hence lead to the violation of three-fold degeneracy in the mass spectrum.
It should be pointed out that our boundary condition \eqref{eq:4.5} is consistent with those obtained by the $\mathbb{Z}_{2}$ orbifold picture \cite{Charmousis:1999rg,Gregory:2000jc,Pilo:2000et,Dubovsky:2003pn,Gherghetta:2005se}, which is consistent with Israel junction condition \cite{Israel:1966rt}, and by the variational principle with the Gibbons-Hawking extrinsic curvature terms \cite{Gibbons:1976ue} at $z=z_{1}$ and $z_{2}$ \cite{Lalak:2001fd,Carena:2005gq,Bao:2005bv}.

Now it is obvious that there is no boundary condition consistent with $\mathcal{N}$ $N=2$ SUSYs with $\mathcal{N}>2$ with the warp factor $A(z) = -\log(z/R)$ on an interval $(z_{1}, z_{2})$.

%%%%% SECTION 5 %%%%%
\section{Conclusion and speculation} \label{sec:5}
In this paper we have studied 5d gravity with the Randall-Sundrum background and showed that two $N=2$ quantum mechanical SUSYs are hidden in the spectrum.
We have also studied boundary conditions consistent with two $N=2$ SUSYs and showed that such boundary condition is uniquely determined.
This result implies that we cannot construct a hierarchy of isospectral Hamiltonians beyond three-fold degeneracy at least in the context of 5d field theory with the Randall-Sundrum background.
Since hierarchy of isospectral Hamiltonians seems very suitable for the structure behind the mass generation mechanism of spin-$N$ particles via compactification, it is very interesting to consider massless higher-spin field theory with extra dimension and study its desired hidden SUSY structure in the spectrum.
Our results shown in section \ref{sec:4} might imply that it would be impossible to construct an infinite Kaluza-Klein tower of massive spin-$N(>2)$ particles with the Randall-Sundrum background.
This is just a speculation, however.
Further studies will be required.

\subsection*{Acknowledgements}
The author would like to thank C. S. Lim, T. Nagasawa, M. Sakamoto, K. Sakamoto and K. Sekiya for collaborations.
This work is supported in part by JSPS Research Fellowships for Young Scientists and JSPS Excellent Young Researcher Overseas Visit Program.

%%%%% REFERENCE %%%%%
\bibliographystyle{utphys}
\bibliography{IFUP-TH-2010-45}

\end{document}